\documentclass[12pt]{iopart}
\usepackage{graphicx}
\usepackage{color}
\eqnobysec

\newcommand{\be}{\[}
\newcommand{\bea}{\begin{eqnarray*}}
\newcommand{\beq}{\begin{equation}}
\newcommand{\beqa}{\begin{eqnarray}}
\newcommand{\ee}{\]}
\newcommand{\eea}{\end{eqnarray*}}
\newcommand{\eeq}{\end{equation}}
\newcommand{\eeqa}{\end{eqnarray}}

\renewcommand{\d}{{\rm d}}

\newcommand{\eq}{{\rm st}}
\newcommand{\eqq}{{\rm eq}}
\newcommand{\rel}{{\rm relax}}

\newcommand{\frad}[2]{\displaystyle{\displaystyle#1\over\displaystyle#2}}

\renewcommand{\i}{{\rm i}}

\newcommand{\la}{\lambda}

\newcommand{\ma}{{\scriptscriptstyle\rm Markov}}
\renewcommand{\max}{{\rm max}}
\newcommand{\mean}[1]{\langle#1\rangle}

\renewcommand{\min}{{\rm min}}
\newcommand{\noneq}{{\rm non-st}}

\newcommand{\s}{{\sigma}}
\newcommand{\sg}{{\rm sg}}
\newcommand{\var}{\mathop{\rm Var}\nolimits}
\newcommand{\C}{{\cal C}}
\newcommand{\D}{\Delta}

\newcommand{\F}{{\cal F}}
\newcommand{\Int}{\mathop{\rm Int}\nolimits}
\renewcommand{\L}{{\rm L}}

\renewcommand{\P}{{\cal P}}
\newcommand{\R}{{\rm R}}
\def\binom#1#2{{#1\choose #2}}
\newcommand{\ens}{{\left\{N_i\right\}}}

\newcommand{\g}{\gamma}
\newcommand{\dpar}{\partial}
\newcommand{\lam}{\lambda}
\newcommand{\fd}{fluc\-tu\-a\-tion-dis\-si\-pa\-tion }

\newcommand{\X}{{\cal X}}

\begin{document}
\title{From urn models to zero-range processes: statics and dynamics}
\author{C Godr\`{e}che\dag\ddag}
\address{\dag\ 
Isaac Newton Institute for Mathematical Sciences,
20 Clarkson Road, Cambridge, CB3 0EH, U.K.}

\address{\ddag\ Service de Physique de l'\'Etat Condens\'e,
CEA Saclay, 91191 Gif-sur-Yvette cedex, France}


\begin{abstract}
The aim of these lecture notes is a description of the statics and dynamics of
zero-range processes and related models. 
After revisiting some conceptual aspects of the subject, 
emphasis is then put on the study of the class of zero-range processes
for which a condensation transition arises.

\end{abstract}


\eads{\mailto{godreche@dsm-mail.saclay.cea.fr}}

\maketitle

 \section*{Introduction}
\label{intro}

The models discussed in these notes are simplified models of physical reality.
Yet, besides the fact that they
play an important role in the elucidation of conceptual problems of
statistical mechanics and probability theory,
they are instrumental in the understanding of a variety of complex
physical situations.

The aim of these notes is a description of the statics and dynamics of
zero-range processes (ZRP)~\cite{spitz} and of related models. 
We present a review of the subject, coming back on some of its conceptual aspects.
We restrict all discussions to homogeneous models where all sites are equivalent.
Before commencing, we summarise in a few words the main organisation of the text.

In Part I: Statics (sections \ref{1}-\ref{stat2}), 
we first show that ZRP are special members of a class of 
stochastic processes
which have the property that their stationary measures are
known and have a product structure.
The probability of a configuration of the system is given by the Boltzmann formula
for an equilibrium urn model with independent sites.
Reversibility (for symmetric dynamics) and pairwise balance
(for asymmetric dynamics) are inherently related to the structure
of the stationary measure.
Generalisations to multiple-species ZRP
are then addressed.
The properties
of the stationary measure of ZRP leading to a 
phase transition between a fluid phase and a condensed phase
are finally briefly reviewed, as a preparation for the second part of these notes.

The stochastic nature of ZRP is fully revealed by the study of their dynamics.
This is the subject of Part II (sections \ref{dyn1}-\ref{dyn3}).
We  first address the nonstationary dynamical behaviour of the system 
when it evolves from a random initial disordered configuration to its stationary state.
Then we investigate some aspects of its stationary dynamics,
when the system fluctuates in its stationary state.
In both cases the model used is that giving rise to condensation.

\bigskip
\section*{ {\large Part I: Statics (sections \ref{1}-\ref{stat2})}}

\section{Dynamical urn models and zero-range processes}
\label{1}
\label{dum}

\label{from}

\subsection{Dynamical urn models}

We name dynamical urn model (DUM) the following stochastic process.
Consider a finite connected graph, made of~$M$ sites (or urns),
on which $N$ particles are distributed.
The occupation $N_i(t)$ of site~$i$ ($i=1,\dots,M$)
is a random variable, and the total number of particles
\be
\sum_{i=1}^M N_i(t)=N
\ee
is conserved in time.
The model is defined by dynamical rules
describing how particles hop from site to site.
An elementary step of the dynamics consists in choosing a departure site~$d$
and an arrival site~$a$ connected to site~$d$,
and in transferring one of the particles present on site~$d$ to site~$a$.
This process takes place with rate $W_{k,l}$ per unit time, 
depending on the occupations both of the departure site, $k=N_d\ne0$, 
and of the arrival site, $l=N_a$\footnote{Throughout this text we use the notation
$N_i$ for the random occupation of site $i$, and $k$ (an integer) for the value
taken by this random variable.}.

On the complete graph (i.e., in the mean-field geometry), all sites are connected,
i.e., sites~$d$ and $a$ are chosen independently at random.
On finite-dimensional lattices, site~$a$ is chosen among the first neighbours
of site~$d$.
In one dimension, site~$a$ is chosen 
to be the right neighbour of site~$d$ with probability $p$,
or its left neighbour with probability $q=1-p$.
In the following we consider the one-dimensional symmetric dynamics,
corresponding to $p=1/2$, and the general asymmetric one,
corresponding to $p\ne 1/2$, both with periodic boundary conditions.

A configuration of the system is specified by the occupation numbers
$N_{i}(t)$, i.e., a complete knowledge of its dynamics involves the
determination of $\mathcal{P}(N_{1,}N_{2},\ldots ,N_{M})$, the probability
of finding the system in a given configuration at time $t$. 

The process can be pictorially viewed in terms of colonies and migration.
The sites are the colonies, or cities. An individual
leaves its colony for another one, with a rate $W_{k,l}$ which depends on the number
of members present in both the departure and the arrival colonies. 
Thus, for example, the philanthrope is characterized by a rate decreasing with $k$ and increasing
with $l$, the misanthrope by the converse.

\subsection{Zero-range processes}
\label{zrp}

Zero-range processes are just particular cases of DUM, 
with the additional restriction that the rate $W_{k,l}$ only depends
on the occupation of the departure site: 
\be
W_{k,l}=u_k.
\ee

This simple restriction is enough to lead to a remarkable property 
of the stationary probability~\cite{spitz,andj}.
Indeed, the probability of a configuration of the system is equal to
\beq
\P(N_{1},\dots,N_{M})=\frac{1}{Z_{M,N}}\prod_{i=1}^M p_{N_i},
\label{product}
\eeq
where it is understood that $\sum N_i=N$, and
where the factor $p_k=p_{N_i=k}$ satisfies the relation
\beq\label{reczrp}
u_k \,p_k=p_{k-1},
\eeq
which leads to the explicit form 
\beq
p_0=1,\qquad p_k=\frac{1}{u_1\dots u_k}.
\label{pk}
\eeq
The normalisation factor, hereafter refered to as the partition function, reads
\beq\label{part}
Z_{M,N}=\sum_{N_{1}}\cdots\sum_{N_{M}}\,p_{N_{1}}\cdots p_{N_{M}}\;\delta
\left(\sum_{i}N_{i},N\right).
\eeq
One important observation to make is that the stationary measure is insensitive 
to the bias.

These results can be proved by inspection.
The master equation at stationarity reads
\beq\label{mastergene}
0=\sum_{\C'\ne \C}M(\C|\C')\,\P(\C')-\sum_{\C''\ne \C}M(\C''|\C)\,\P(\C),
\eeq
where
$\C=\{N_1,\ldots ,N_M \}$,
and $M(\C|\C')$ is the transition rate from $\C'$ to $\C$.
Consider a system of $M=3$ sites for simplicity.
At stationarity the master equation reads explicitly
\beqa
&p&\left[\P(N_1+1,N_2-1,N_3)\,u_{N_1+1}(1-\delta(N_2,0))+{\rm c.p.}+{\rm c.p.}\right]
\nonumber\\
+&q&\left[\P(N_1+1,N_2,N_3-1)\,u_{N_1+1}(1-\delta(N_3,0))+{\rm c.p.}+{\rm c.p.}\right]
\nonumber\\
=&&
\P(N_1,N_2,N_3)[\,u_{N_1}(1-\delta(N_1,0))+{\rm c.p.}+{\rm c.p.}],
\eeqa
where c.p. stands for circular permutation.
Carrying the product form~(\ref{product})
into the equation, and using (\ref{reczrp}), satisfies the master equation.
This is the {\it unique} solution of the problem.

The cancelation of terms in the equation occurs by pair.
Pairs correspond to terms bearing the same $p$ (respectively $q$) factor,
and the same $1-\delta(N_i,0)$ factor.
Hence we have for example
\beq\label{dba}
p\P(N_1+1,N_2-1,N_3)\,u_{N_1+1}=p\P(N_1,N_2,N_3)\,u_{N_2},
\eeq
i.e., using the product form~(\ref{product}), with $N_1=k$ and $N_2=l$,
\beq\label{dbb}
p_{k+1}\,p_{l-1}\,u_{k+1}=p_k\, p_l\, u_l,
\eeq
which is precisely the relation that leads to~(\ref{reczrp}).

It is interesting to emphasize the interpretation of~(\ref{dba}), or (\ref{dbb}).
Consider the following configurations:
\bea
\C&=&(N_1,N_2,N_3),\\
\C'&=&(N_1+1,N_2-1,N_3),\\
\C''&=&(N_1,N_2-1,N_3+1),
\eea
and the corresponding rates
\bea
M(\C|\C')=p\,u_{N_1+1},\\
M(\C'|\C)=q\,u_{N_2},\\
M(\C''|\C)=p\,u_{N_2}.
\eea
In general, i.e. for a general value of $0<p<1$, (\ref{dba}) reads
\be
M(\C|\C')\P(\C')=M(\C''|\C)\P(\C).
\ee
This is a condition for {\it pairwise balance}~\cite{pairwise}.
It expresses the equality between the probability fluxes flowing from 
 $\C'$ to $\C$, and from $\C$ to $\C''$.
In the particular case where $p=1/2$, or more generally when the
dynamics is symmetric, then (\ref{dba}) becomes the condition for
detailed balance
\be
M(\C|\C')\P(\C')=M(\C'|\C)\P(\C).
\ee
   

\subsection{Equilibrium urn models with independent sites}
\label{equrn}

We now adopt a completely different point of view.
We consider equilibrium urn models with independent sites, on which
a dynamics is then defined, in such a way that equilibrium
is recovered at long times.

As above, we consider a finite connected graph, made of~$M$ sites (or urns),
on which $N$ particles are distributed.
The number of particles on site $i$ is the random variable $N_i$,
with $\sum N_i=N$.
The total energy of the system is defined as the sum
\[
E(N_{1},\dots,N_{M})=\sum_{i=1}^{M}E(N_{i}).
\]
Let
\beq\label{pni}
p_{N_{i}}=\e^{-\beta E(N_i)}
\eeq
be the unnormalized Boltzmann weight attached to site $i$.
Then, clearly, the probability of a configuration of the system is
given by the product form~(\ref{product}), and $Z_{M,N}$ 
appears as the usual partition function for this statistical mechanical
system.

We now define a dynamics for this model, such
that equilibrium is attained in the limit of long times.
We therefore choose a rule obeying detailed balance for the move of a particle.
This implies that the dynamics should be symmetric.
Restricting to the one-dimensional case,  ($p=1/2$),
if $N_i=k$ and $N_{i\pm1}=l$, we have
\beq\label{db1}
p_k p_l\,W_{k,l}=p_{k-1}p_{l+1}\,W_{l+1,k-1},
\eeq
which expresses the probability balance between the configurations
$\{N_d=k,N_a=l\}$ where ($d=i, a=i\pm1$), and $\{N_d=l+1,N_a=k-1\}$ 
where ($d=i\pm1,a=i$).
It applies as well to the case of the complete graph.
For example, with the Metropolis rule, the move is
allowed with probability $\min(1,\exp(-\beta\Delta E))$,
where $\Delta E$ is the change in energy due to the move.

Let us mention two well-studied models in this class: the backgammon
model and the zeta-urn model, that we briefly describe.
The backgammon model is a simple example of a system
which exhibits slow relaxation due to entropy barriers~\cite{ritort,barcurn}.
The following choice of an energy function is done:
\be
E(N_i)=-\delta(N_i,0).
\ee
The statics of this model is trivial.
Its interest  lies in its dynamical behaviour.
The dynamics of the model has been thoroughly studied in the mean-field geometry,
with Metropolis dynamics, and with the additional rule 
that a particle (instead of a site) is chosen at random.
The rate for the Metropolis rule reads
\be
W_{k,l}=\min \left (1,\frac{p_{k-1}p_{l+1}}{p_kp_l} \right ).
\ee
From (\ref{pni}), we have
\be
p_0=\e^\beta,\qquad p_k=1, \quad (k>1),
\ee
and therefore $W_{k,0}=\e^{-\beta}$ for any $k>1$, and $W_{k,l}=1$ otherwise,
or in compact form:
\be
W_{k,l}=1+(\e^{-\beta}-1)\delta_{l,0} (1-\delta_{k,1})
\qquad (k>0).
\ee
As can be read on this expression,
at low temperature increasing the number of empty sites is not favoured.
The total energy is indeed equal
to minus the number of empty sites,
so that particles tend to condensate in fewer and fewer sites as times passes,
at least at low temperature.

The static zeta urn model has 
energy function
\beq
E(N_i)=\ln(N_i+1),
\label{ze}
\eeq
hence 
\be
p_k=\frac{1}{(1+k)^{\beta}}.
\ee
The model was initially introduced
as a mean-field model of discretized quantum gravity~\cite{bia}.
Its dynamics was subsequently defined and investigated in the mean-field geometry
with heat-bath dynamics~\cite{zeta1,zeta2}.

If instead, the transfer rate is taken to be that of a ZRP,
with $W_{k,l}=u_k$, where $u_k=p_{k-1}/p_k$, the
universal properties of the dynamics of the zeta urn model are not changed~\cite{cg}. 
We thus get
\be
u_k=\left ( 1+\frac{1}{k}\right)^\beta\approx 1+\frac{\beta}{k}.
\ee
The model is therefore in the same universality class as
the ZRP with condensation
studied in the rest of this text, and defined with the rate
$u_k=1+b/k$.
The parameter $b$ for this model
can  therefore be identified with the inverse temperature.

To summarise at this point, we have so far encountered two classes of dynamical urn models
with stationary product measures. On the one hand, ZRP are 
defined for any value of
the drive, and are such that the transfer rate
$W_{k,l}$ only depends on $k$.
On the other hand, equilibrium urn models with independent sites
are defined from the start without drive, but the transfer rate 
has the full dependence in both $k$ and $l$.
A natural question to ask is whether there exist models possessing
both features, namely models with stationary
product measure, even when submitted to a drive, and with transfer
rate $W_{k,l}$ not restricted to depend only on $k$.

\subsection{Dynamical urn models with stationary product measure}

We address the question just posed.
Given a DUM,
what choice of rate $W_{k,l}$ is compatible with 
a stationary measure of the form (\ref{product}),
{\em even if the dynamics is not symmetric}?\\
Let us restrict to 
the case of the one-dimensional geometry with asymmetric hops.
The results are as follows:

\begin{itemize}
\item
In the general case, $0<p<1$,  two conditions are imposed on the rate $W_{k,l}$. 
The first condition is 
\beq\label{db}
p_k p_l\,W_{k,l}
=p_{k-1} p_{l+1}\,W_{l+1,k-1}.
\eeq
The second condition reads
\beq\label{constraint}
W_{k,l}-W_{l,k}=W_{k,0}-W_{l,0}.
\eeq
Eq.~(\ref{db})  
expresses the condition of pairwise balance.

\item
In the symmetric case ($p=1/2$), the only 
condition imposed on the transfer rate is (\ref{db}), or equivalently~(\ref{db1}).
It expresses the condition of detailed balance.
In other words, if the stationary measure is a product,
it is necessarily an equilibrium measure and we are taken back to the 
situation of section~\ref{equrn}.

\end{itemize}

Let us give the proof.
By hypothesis, the stationary probability is given 
and has the product form (\ref{product}), with given $p_{N_i}$.
We rewrite the master equation (\ref{mastergene})
as an equality between gain and loss terms, after dividing both hand sides by $\P(\C)$,
\be
p G_R+q G_L=p L_R+q L_L,
\ee
with right and left contributions
\bea
L_R&=&\sum_i W_{N_i,N_{i+1}},\qquad L_L=\sum_i W_{N_{i+1},N_{i}},\\
G_R&=&\sum_i W_{N_i+1,N_{i+1}-1}\,\frac{p_{N_i+1}p_{N_{i+1}-1}}{p_{N_i}p_{N_{i+1}}},\\
G_R&=&\sum_i W_{N_{i+1}+1,N_{i}-1}\,\frac{p_{N_{i+1}+1}p_{N_i-1}}{p_{N_i}p_{N_{i+1}}} .
\eea

We now specialize to the configuration where all sites are empty except for sites $i$
and $i+1$:
\be
\C=\{N_1=0,\ldots,N_{i-1}=0,N_i=k,N_{i+1}=l,N_{i+2}=0,\ldots,N_M=0\}.
\ee
We obtain
\beqa\label{bilan}
p \left( W_{1,k-1}\,\frac{p_1p_{k-1}}{p_0 p_k}
+W_{k+1,l-1}\,\frac{p_{k+1} p_{l-1}}{p_k p_l}\right)\nonumber\\
+q\left(W_{1,l-1}\,\frac{p_1p_{l-1}}{p_0 p_l}
+W_{l+1,k-1}\,\frac{p_{k-1} p_{l+1}}{p_k p_l}\right)\nonumber
\\=
p(W_{k,l}+W_{l,0})+q(W_{l,k}+W_{k,0}).
\eeqa
Taking $k=0$, (\ref{bilan}) reduces to
\beq\label{db0}
{p_1p_{l-1}}\,W_{1,l-1}
=
{p_0 p_l}\,W_{l,0},
\eeq
which expresses the probability balance between the configurations 
$\{N_d=1,N_a=l-1\}$ and $\{N_d=l,N_a=0\}$.
This equality is then used in (\ref{bilan}) to yield
the fundamental equation
\beqa\label{fund}
p(W_{k,l}-W_{k,0})+q(W_{l,k}-W_{l,0})=\nonumber\\
p\, \left(W_{k+1,l-1}\,\frac{p_{k+1} p_{l-1}}{p_k p_l}-W_{l,0}\right)
+q\,\left( W_{l+1,k-1}\,\frac{p_{k-1} p_{l+1}}{p_k p_l}-W_{k,0}\right).
\eeqa
From this equation, the two conditions (\ref{db})  and (\ref{constraint}) are 
obtained, as shown in the appendix.
The conditions thus found are necessary. 
They are also sufficient as one can convince oneself by redoing the
reasoning for a generic configuration.
The analysis done here applies as well to the complete graph, 
for which the dynamics is symmetric.

Coming back to the case of a ZRP,
condition~(\ref{constraint}) is trivially satisfied, while the 
pairwise balance condition (\ref{db}) yields (\ref{dbb}), rewritten here for convenience,
\be
p_kp_l\,u_k=p_{k-1}p_{l+1}\,u_{l+1}.
\ee
The ZRP appears as the minimal model of the class of DUM 
leading to a product measure in the stationary state independent of the asymmetry.
It is important to realize that this measure is 
that of an equilibrium urn model with independent sites (see section \ref{equrn})
and therefore any result on the statics of a ZRP
pertains to the field of equilibrium statistical mechanics.

The original work on the question posed in the present section
is due to \cite{coco}.
The  dynamical urn model described in the present notes
is named a {\it misanthrope process} in~\cite{coco}  because
the rates
 $W_{k,l}$ considered in this reference are increasing functions of $k$.
Yet another presentation, restricted to the 1D totally asymmetric case ($p=1$)
can be found in~\cite{evans2}.

\section{A counterexample}
\label{v}

Let us now examine the case where the transfer rate only depends on the
occupation of the arrival site, 
\beq\label{vl}
W_{k,l}=v_l (1-\delta_{k,0}).
\eeq
If the dynamics is symmetric, the only constraint to take into
account in order to have product probability 
in the stationary state is the detailed balance condition~(\ref{db}),
which reads here
$p_{k+1}p_{l-1}\,v_{l-1}=p_k p_l\,v_k$.
The relation $p_{l-1}\,v_{l-1}=p_l$ follows, which determines
the measure fully.
However, if the dynamics is not symmetric,
(\ref{constraint}) is violated by (\ref{vl}),
which rules out the possibility of stationary product measure for this case. 

Let us illustrate the difficulty
on the simple case of a system of $M=3$ sites.
The stationary master equation reads
\bea
&p&\left[\P(N_1+1,N_2-1,N_3)\,v_{N_2-1}(1-\delta(N_2,0))+{\rm c.p.}+{\rm c.p.}\right]
\nonumber\\
+&q&\left[\P(N_1+1,N_2,N_3-1)\,v_{N_3-1}(1-\delta(N_3,0))+{\rm c.p.}+{\rm c.p.}\right]
\nonumber\\
=
&p&\P(N_1,N_2,N_3)[v_{N_1}(1-\delta(N_3,0))+{\rm c.p.}+{\rm c.p.}]\\
+&q&\P(N_1,N_2,N_3)[v_{N_1}(1-\delta(N_2,0))+{\rm c.p.}+{\rm c.p.}].
\eea
There is no way of pairing
the terms in the master equation to obtain their mutual cancellation
if $p\ne 1/2$, while this is possible for $p=1/2$.
More generally,
the stationary probability is unknown for the asymmetric process (for arbitrary
system size $M$)~\cite{ustocome}.

\section{Two-species ZRP: conditions for product measure}
\label{zrp2}

A simple generalization of the ZRP defined so far consists
in considering  two (or more generally $n$)  coexisting species
on each site~\cite{gross2,hanney}, named 
particles of type A and B respectively.
The hopping rates for A and B particles only depend on the occupations of the
departure site: $N_i^{A}=k$, $N_i^{B}=l$.
They are respectively
denoted by $u_{k,l}$ and $v_{k,l}$. 
The new fact is that the condition for product stationary measure
imposes a constraint on the rates $u_{k,l}$ and $v_{k,l}$~\cite{gross2,hanney},
given by equation~(\ref{contrainte2}).


We revisit this problem,
keeping the line of thought followed for the (single-species) ZRP in section~\ref{from}.
We want to show that, as was the case for the single-species ZRP, 
for the two-species ZRP satisfying~(\ref{contrainte2}) the following properties come together:
\begin{itemize}
\item
The stationary probability is a product and is insensitive to the bias.
It is the stationary probability of an equilibrium urn model with independent sites.
\item
If the dynamics is symmetric, the process is reversible, i.e. satisfies
detailed balance, otherwise, in the presence of a bias, pairwise balance 
holds.
\end{itemize}

\subsection{Equilibrium urn models with independent sites}

Let us first consider an equilibrium urn model for two species
with independent sites.
A configuration of the system is denoted by 
$\C=\{\vec {N_1},\ldots,\vec{N}_M\}$,
where $\vec {N_i}=(N_i^{A},N_i^{B})$.
The energy is given by the sum
\be
E(\C)=\sum_iE(\vec {N_i}).
\ee
The Boltzmann weight reads
\beq\label{product2}
\P(\C)=\frac{1}{Z_{M,N^{A},N^{B}}}\prod_i p_{\vec {N}_i},
\eeq
where $N^{A}$ and $N^{B}$ are respectively the total number of A and B particles,
and $Z$ the partition function.
A dynamics yielding this equilibrium measure should fulfill detailed balance.
We restrict the rates to depend only on the departure site.
With for the departure site: $N_d^A=k,N_d^B=l$, and the arrival site:
$N_a^A=m,N_a^B=n$, we must impose
\beqa\label{dbzrp2}
p_{k,l}\,p_{m,n}\,u_{k,l}=p_{k-1,l}\,p_{m+1,n}\,u_{m+1,n}\nonumber\\
p_{k,l}\,p_{m,n}\,v_{k,l}=p_{k,l-1}\,p_{m,n+1}\,v_{m,n+1},
\eeqa
hence, 
\beq\label{rec2}
u_{k,l}\,p_{k,l}=p_{k-1,l},\qquad v_{k,l}\,p_{k,l}=p_{k,l-1}.
\eeq
These relations generalise (\ref{reczrp}).
Consideration of the two possible paths leading from $p_{k,l}$ to $p_{k-1,l-1}$,
using~(\ref{contrainte2}),
imposes a ``gauge" condition on the rates:
\beq\label{contrainte2}
u_{k,l}\,v_{k-1,l}=v_{k,l}\,u_{k,l-1}.
\eeq

\subsection{Product measure}
We can now proceed as for the single-species ZRP. 
We claim that, even in the presence of a bias, 
(\ref{product2}), (\ref{rec2}), hold.
The proof is by inspection:
(\ref{product2}) carried into the master equation of the process
is seen to be a solution if (\ref{dbzrp2}), or (\ref{rec2}), hold.
The constraint (\ref{contrainte2}) follows.
See \cite{gross2,hanney}.


\subsection{ Reversibility implies stationary product measure}

Finally we show by a direct route that (\ref{contrainte2}) is a consequence of
reversibility,
when the dynamics is symmetric. 
We use the Kolmogorov condition, a necessary and sufficient condition for
the reversibility of a Markov process (i.e., for detailed balance to hold),
which states that the product of rates along any cycle in the state space
of the process and for the reverse cycle should be equal~\cite{kol,kel}.
Consider the configuration
\be
(\vec{N}_{1}=(k,l), \vec{N}_2=(m,n),\ldots).
\ee
We consider the following cycle in the space of states of the process
\bea
(k,l;m,n)\rightarrow (k,l-1;m,n+1)\rightarrow 
(k+1,l-1;m-1,n+1)\\\rightarrow (k+1,l;m-1,n)\rightarrow (k,l;m,n).
\eea
For the cycle considered above, the Kolmogorov condition yields
\be
\frac{v_{k,l}\,u_{k+1,l}}{v_{k+1,l}\,u_{k+1,l-1}}
=\frac{u_{m,n}\,v_{m,n+1}}{u_{m,n+1}\,v_{m-1,n+1}}.
\ee
This condition is satisfied if and only if (\ref{contrainte2}) holds.

\subsection{An example of a two-species ZRP with non product stationary measure}

Consider the ZRP defined by the following rates~\cite{glm}
\be
u_{k,l}=1+\frac{b}{l},\qquad v_{k,l}=1+\frac{b}{k}.
\ee
These rates violate (\ref{contrainte2}), and therefore, as explained above,
this process
violates time reversal symmetry {\it even in the absence of a bias}.
The study of the stationary properties of the model is addressed 
in~\cite{cg2}.

\section{Two extreme cases}
\label{2sit}

\subsection{The case of two sites}
We come back to the case of a general dynamical urn model, 
with one species, where now the number of sites is $M=2$.
This case is interesting for several reasons.
Firstly the model stands by itself, for instance the Ehrenfest urn model belongs
to this class, as shown below.
Secondly, it illustrates some aspects of the general theory for a system of arbitrary size $M$.
Finally, it relates to the other case considered in this section,
a thermodynamic system on the complete graph, the master equation of which
is formally that of a two-site system.

Since $N_2=N-N_1$, 
a configuration of the system is entirely defined by the occupation of site 1, 
$N_1$, and the hopping rate only depends on one variable: $W_{k,l}=u_k$.
Let us denote the occupation probability of site 1, i.e., 
the probability of a configuration of the system, by 
\be
f_k(t)=\P(N_1(t)=k).
\ee
It obeys the master equation
\beqa\label{2sites}
\frac{\d f_k(t)}{\d t} &=&\mu_{k+1}\,f_{k+1}+\la_{k-1}\,f_{k-1}-(\mu
_k+\la_k)f_k\qquad(1\le k\le N-1),\nonumber\\
\frac{\d f_0(t)}{\d t} &=&\mu_1\,f_1-\la_0f_0,\\
\frac{\d f_N(t)}{\d t} &=&\la_{N-1}\,f_{N-1}-\mu_Nf_N,\nonumber
\eeqa
where $\la_k$ and $\mu_k$
are respectively the rate at which a particle enters site~1, coming
from site~2, or leaves site~1 for site~2:
\be
\la_k=u_{N-k},\qquad \mu_k=u_k.
\ee
The equations for $k=0$ or $k=N$ are special, since $u_0=0$.
The above equations describe a biased random walk on the interval $(0,N)$,
with reflecting boundaries at 0 and $N$,
the position of the walker being the random variable $N_1(t)$,
i.e., the number of particles on site~1.

The time-independent solution to~(\ref{2sites}) 
satisfies
\be
\mu_{k+1}\,f_{k+1,\eqq}-\lambda_k\,f_{k,\eqq}=
\ldots=
\mu_{1}\,f_{1,\eqq}-\lambda_0\,f_{0,\eqq}=0,
\ee
which yields the detailed balance condition at equilibrium
\beq\label{db2}
\mu_{k+1}\,f_{k+1,\eqq}=\lambda_k\,f_{k,\eqq}.
\eeq
From this equation it is easy to obtain
\be
f_{k,\eqq}=\frac{p_k\,p_{N-k}}{Z_{2,N}},\qquad Z_{2,N}=\sum_{k=0}^N p_kp_{N-k},
\ee
where the $p_k$ are given by (\ref{pk}).
These expressions are special instances of eqs.~(\ref{product}) and (\ref{part}) which hold for the general case.
Elements on the dynamics of the two-site model can be found in~\cite{gl2005}.

\medskip
\noindent
{\it Remark.\ }
This process is equivalent to the historical
Ehrenfest model~\cite{ehr,kac}, defined
as follows.
Consider $N$ particles, labeled from 1 to $N$, which
are distributed in two urns (sites).
At random times, given by a Poisson process with unit rate,
a particle is chosen at random
(i.e., an integer between 1 and~$N$ is chosen at random),
and moved from the site on which it is to the other site.
The master equation reads
\beq
\frac{\d f_k(t)}{\d t}=\frac{k+1}{N}f_{k+1}(t)+\frac{N+1-k}{N}f_{k-1}(t)
-f_k(t).
\label{df}
\eeq
Indeed, a move of a particle from site number~1 to site number~2
(resp.~from site number~2 to site number~1) 
occurs with a rate $k/N$ (resp.~$(N-k)/N$) per unit time.

Note that the rule of choosing a labeled particle is different from the rule
adopted above for dynamical urn models
(as was already the case for the backgammon model).
Yet we can describe this model as a 2-site dynamical urn model, by taking
$u_k=k$ (dropping the factor $N$ which enters the scale of time). 
Then $p_k=1/k!$, and the distribution of particles
amongst the two sites is binomial,
\beq
f_{k,\eqq}=2^{-N}\binom{N}{k}\quad(k=0,\dots,N),
\eeq
as is well known for the Ehrenfest model.

\subsection{A thermodynamic system on the complete graph}
In the mean-field geometry, for a thermodynamic system, the temporal evolution of the occupation
probability $f_{k}(t)$ is given by the master equation
\begin{eqnarray}\label{master}
\frac{\d f_{k}(t)}{\d t} &=&\mu _{k+1}\,f_{k+1}+\lambda _{k-1}\,f_{k-1}-(\mu
_{k}+\lambda _{k})f_{k}\qquad (k\geq 1), \nonumber\\
\frac{\d f_{0}(t)}{\d t} &=&\mu _{1}\,f_{1}-\lambda _{0}f_{0},
\end{eqnarray}
where
\begin{equation}\label{rates}
\mu _{k}=u_k,\quad (k>0),
\qquad \lambda _{k}=\sum_{l=1}^{\infty }u_lf_{l}\equiv \bar{%
u}_{t},\quad (k\ge0). 
\end{equation}
These are respectively the rates at which a particle leaves site 1, or arrives
on this site.
In other words, on the complete graph, all sites
other than site 1 play the role of a single site 
from which particles are emitted with rate $\bar u_t$,
and therefore (\ref{master}) is formally similar to the master equation~(\ref{2sites})
for a system of two sites.
In the present case
this set of equations is non linear because $\bar{u}_{t}$
is itself a function of the $f_{k}(t)$. 

In the stationary state
the detailed balance condition~(\ref{db2}) reads
\be
\frac{f_{k+1,\eqq}}{f_{k,\eqq}}=\frac{\lambda _{k}}{\mu _{k+1}}
=\frac{\bar u_\eqq}{u_{k+1}},
\ee
yielding
\be
f_{k,\eqq}=\frac{\lambda _{0}\ldots \lambda _{k-1}}{\mu _{1}\ldots \mu _{k}}%
f_{0,\eqq},
\ee
where $f_{0,\eqq}$ is fixed by normalisation.
Hence
\begin{equation}\label{fkmf}
f_{k,\eqq}=\frac{p_{k}\bar u_\eqq^{k}}{\sum_{k=0}^{\infty }p_{k}\bar u_\eqq^{k}},
\end{equation}
with the $p_k$ given by (\ref{pk}).
This expression is a particular instance of the general case (\ref{fkst}).

\section{Statics of ZRP: Fundamental properties}
\label{stat1}

We collect here the results found so far concerning single-species ZRP's.
A ZRP is a dynamical urn model, for which the rate of transfer of a particle, $u_k$,
only depends on the occupation of the departure site, $k$.
The stationary state of a ZRP is that of an equilibrium urn model with independent sites:
the probability of a configuration of the system is
(independently of the asymmetry)
\beq\label{z:factor}
\P(N_{1},\dots,N_{M})=\frac{1}{Z_{M,N}}\prod_{i=1}^M p_{N_i},
\eeq
with partition function
\beq\label{z:part}
Z_{M,N}=\sum_{N_{1}}\cdots\sum_{N_{M}}\,p_{N_{1}}\cdots p_{N_{M}}\;\delta
\left(\sum_{i}N_{i},N\right).
\eeq
The factor $p_{N_i}$ obeys the
pairwise balance condition (\ref{db}), i.e., $p_kp_l\,u_k=p_{k-1}p_{l+1}\,u_{l+1}$, 
and hence
\be
p_k\,u_k=p_{k-1},
\ee
which gives the explicit form of $p_k$ (for $u_k$ given)
\beq
p_0=1,\qquad p_k=\frac{1}{u_1\dots u_k}.
\label{z:pk}
\eeq
The value given to $p_0$ is arbitrary.
The energy function associated to the underlying equilibrium urn model
mentioned above is defined using
eq.~(\ref{pni}).


The partition function $Z_{M,N}$ 
obeys the recursion formula
\beq\label{zrecrel}
Z_{M,N}=\sum_{k=0}^N p_k\,Z_{M-1,N-k}.
\eeq
This ensures that the stationary single-site occupation probability
\beq\label{fkeq}
f_{k,\eq}=\P(N_1=k)=\frac{p_k\,Z_{M-1,N-k}}{Z_{M,N}}
\eeq
is normalised.
We have
\beq\label{z0z1z2}
Z_{0,N}=\delta_{N,0},\qquad
Z_{1,N}=p_N,\qquad
Z_{2,N}=\sum_{k=0}^N p_kp_{N-k},
\eeq
and so on.
Using an integral representation of the Kronecker delta function, 
\be
\delta(m,n)=\oint\frac{\d z}{2\pi\i z^{n+1}}\,z^m,
\ee
we obtain
\beq
Z_{M,N}=\oint\frac{\d z}{2\pi\i z^{N+1}}\,P(z)^M,
\label{contour}
\eeq
where the generating series of the weights $p_k$ reads
\be
P(z)=\sum_{k\ge0} p_kz^k.
\ee
In other words, $Z_{M,N}$ is the coefficient of $z^N$ in $P(z)^M$.
Static properties of the ZRP are therefore entirely encoded in this series.

In the thermodynamic limit ($M\to\infty$ at fixed density $N/M=\rho$),
the free energy per site,
\be
\F=-\lim_{M\to\infty}\frac{1}{M}\ln Z_{M,N},
\ee
can be obtained by evaluating the
contour integral in~(\ref{contour}) by the saddle-point method.
The saddle-point value $z_0$ depends on the density $\rho$ through the equation
\beq
\frac{z_0 P'(z_0)}{P(z_0)}=\rho.
\label{col}
\eeq
The free energy per site is
$\F=\rho\ln z_0-\ln P(z_0)$, 
and the stationary occupation probability reads
\beq\label{fkst}
f_{k,\eq}=\frac{p_k\,z_0^k}{P(z_0)}.
\eeq
Eq. (\ref{col}) can be rewritten as
\beq\label{col2}
\langle N_1\rangle=\sum_k k f_{k,\eq}=\rho.
\eeq
Note that the function 
\be
\rho(z_0)= z_0\frac{P'(z_0)}{P(z_0)}
\ee
is increasing with $z_0$ because 
\be
z_0\frac{\d\rho(z_0)}{\d z_0}=\var N_1.
\ee

Finally the stationary average rate reads
\beq\label{ubarstat}
\bar u_\eq(M,N)=\langle u_{N_1}\rangle=\sum_k u_k\,f_{k,\eq}
=\sum_k u_k\,
\frac{p_k\,Z_{M-1,N-k}}{Z_{M,N}}
=\frac{Z_{M,N-1}}{Z_{M,N}}.
\eeq
In the thermodynamic limit, we have $\bar u_\eq=z_0$ (defined in (\ref{col}) above).
The expression (\ref{fkmf}) found for the case of the complete graph in the
thermodynamic limit is a particular example of (\ref{fkst}).

\section{Statics of ZRP: Examples and the phenomenon of condensation}
\label{stat2}

We illustrate through examples the considerations of the previous section.
In particular we discuss the possible solutions of eq. (\ref{col}) (or (\ref{col2})).
Two possible situations can arise. Either 
$\rho(z_0)$ is allowed to increase without bounds, in which case the equation has a solution
in $z_0$ for any value of $\rho$. Or $\rho(z_0)$ reaches a maximal value, $\rho_c$, in which case
the equation has no solution if $\rho>\rho_c$.

\subsection{Two simple examples }

Let $u_k=k$.
This model can be seen as a 
multi-urn generalisation of the Ehrenfest model.
We have $P(z)=\e^z$.
The radius of convergence of this series is infinite.
Hence eq. (\ref{col}) has a solution for any value of $\rho$:
$\rho(z_0)=z_0$, hence $z_0=\rho$, and 
\be
f_{k,\eq}=\e^{-\rho}\frac{\rho^k}{k!},
\ee
which is a Poisson distribution. 
The fast decay of the distribution is characteristic of 
an homogeneous fluid phase.

As a second example let $u_k=1$.
Then $P(z)=1/(1-z)$.
The partition function of a finite system is
\be
Z_{M,N}=\binom{M+N-1}{N}.
\ee
The radius of convergence of $P(z)$ is equal to 1.
At this maximal allowed value of $z$, $\rho(z)=z/(1-z)$ is infinite.
Therefore (\ref{col}) has a solution for any value of $\rho$:
$z_0=\rho/(1+\rho)$,
and finally 
\be
f_{k,\eq}=\frac{1}{1+\rho}\left(\frac{\rho}{1+\rho}\right)^k.
\ee
The system is again in a fluid phase.

\subsection{The canonical example for the phenomenon of condensation}
\label{canon}

We consider the ZRP with transfer rate
\be
u_k=1+\frac{b}{k}.
\ee
This case, and closely related models, have been studied in various 
references~\cite{bia,evans1,loan,maj,cg,gross,zeta1,zeta2,wis1,harris}.
We follow here the approach and notations of~\cite{gl2005}.
For this choice of rate, 
\beqa\label{pkas}
&&p_k=\frac{\Gamma(b+1)\,k!}{\Gamma(k+b+1)}
=\int_0^1 \d u\,u^k\,b(1-u)^{b-1}\approx\frac{\Gamma(b+1)}{k^b},\nonumber\\
&&P(z)=\int_0^1\d u\,\frac{b(1-u)^{b-1}}{1-zu}
=\null_2F_1(1,1;b+1;z),
\eeqa
where $_2F_1$ is the hypergeometric function.
The function $P(z)$ has a branch cut at $z=1$, with a singular part of the 
form\footnote{Whenever $b=n\ge2$ is an integer, the amplitude $A$ diverges.
The singular part of the generating series is of the form
$P_\sg(z)\approx n(-1)^n(1-z)^{n-1}\ln(1-z)$.}
\be
P_\sg(z)\approx A\,P(1)(1-z)^{b-1},
\ee
so that $P(z)$ is only differentiable $n\equiv\Int(b)-1$ many times at $z=1$:

\be
P(z)\approx P(1)+(1-z)\,P'(1)+\cdots +\frac{(1-z)^n}{n!}P^{(n)}(1)+P_\sg(z).
\ee
The following values are of interest:
\beq
\matrix{
P(1)=\frad{b}{b-1},\hfill&A=\frad{(b-1)\pi}{\sin\pi b},\hfill\cr
P'(1)=\frad{b}{(b-1)(b-2)},\quad&P''(1)=\frad{4b}{(b-1)(b-2)(b-3)}.}
\eeq

For $b\le2$, $\rho(1)$ is infinite. The system is in a fluid phase:
\beq\label{fkflu}
f_{k,\eq}\sim k^{-b}\,\e^{-k |\ln z_0|}
\eeq
 
For $b>2$, $\rho(1)$ is finite. The system has a continuous phase transition
at a finite critical density
\be
\rho_c=\frac{P'(1)}{P(1)}=\frac{\sum_k k\,p_k}{\sum_k p_k}=
\frac{1}{b-2},
\ee
such that the saddle point $z_0$ reaches the singular point $z=1$.
This critical density separates a fluid phase $(\rho<\rho_c)$
and a condensed phase $(\rho>\rho_c)$.

\subsubsection*{Fluid phase $(\rho<\rho_c)$.}
The equation (\ref{col}) has a solution for any $\rho<\rho_c$.
The single site probability has the form (\ref{fkflu}).

\subsubsection*{Critical density $(\rho=\rho_c)$.}

The occupation probability
\beq
f_{k,\eq}=\frac{p_k}{P(1)}\approx\frac{(b-1)\Gamma(b)}{k^b}
\label{fkc}
\eeq
falls off as a power-law in the thermodynamic limit.
The critical free energy reads
\be
\F_c=-\ln P(1)=-\ln\frac{b}{b-1}.
\ee
The second moment of the occupation probability,
\beq\label{second}
\mu_c=\langle N_1^2\rangle =\sum_{k\ge0}k^2\,f_{k,\eq}=\frac{P'(1)+P''(1)}{P(1)}
=\frac{b+1}{(b-2)(b-3)},
\eeq
is convergent for $b>3$ (regime of normal fluctuations),
and divergent for $2<b<3$ (regime of anomalous fluctuations).

\subsubsection*{Condensed phase ($\rho>\rho_c$).}

A large and finite system in the condensed phase
essentially consists of a uniform critical background,
containing on average $N_c=M\rho_c$ particles,
and of a macroscopic condensate, containing on average
$\D=N-N_c=M(\rho-\rho_c)$ excess particles with respect to the critical state.

The occupation probability $f_{k,\eq}$ accordingly splits
into two main contributions~\cite{maj}.
The contribution of the critical background,
corresponding to small values of the occupation ($k\ll M)$,
is approximately given by~(\ref{fkc}).
The contribution of the condensate shows up as a hump located around $k=\D$.
The hump is a Gaussian whose width scales as $M^{1/2}$ whenever $\mu_c$
is finite, i.e., for $b>3$, whereas it has power-law tails and a larger width,
scaling as $M^{1/(b-1)}$, in the regime of anomalous fluctuations $(2<b<3)$.
The weight of the condensate probability hump is approximately $1/M$,
in accord with the picture that the system typically
contains a well-defined condensate located on a single site at any given time.

\subsection{Rate $u_k=1+a/k^\s$: Stretched-exponential critical behaviour}

Consider the ZRP with
hopping rate~\cite{evans1,wis1,gl2005}
\beq
u_k=1+\frac{a}{k^\s},
\label{grates}
\eeq
where $\s$ is an arbitrary exponent.
The situation of interest corresponds to $0<\s<1$.
Equation~(\ref{pk}) leads to the estimate
\beq
p_k\sim\exp\left(-a\sum_{\ell=1}^k\frac{1}{\ell^\s}\right)
\sim\exp\left(-\frac{a}{1-\s}\,k^{1-\s}\right).
\label{gdecay}
\eeq
The generating series $P(z)$ has an essential singularity at $z=1$
with an exponentially small discontinuity.
The critical density
\be
\rho_c=\frac{P'(1)}{P(1)}=\frac{\sum_k k\,p_k}{\sum_k p_k}
\ee
is finite.
The occupation probability at the critical density, $f_{k,\eq}=p_k/P(1)$,
decays as a stretched exponential law.

\bigskip
\section*{ {\large Part II: Dynamics (sections \ref{dyn1}-\ref{dyn3})}}

\section{Zero-range processes: nonstationary dynamics (I)}
\label{dyn1}

The question is to determine the
temporal evolution of the system starting from a random disordered initial condition.
Here we study the dynamics of the class of ZRP
giving rise to a condensation transition in their stationary state.
For simplicity we will choose the hopping rate 
\be
u_k=1+\frac{b}{k}.
\ee
We address the question first in the fully connected geometry.

The same question can be addressed for dynamical urn models 
(see e.g.~\cite{barcurn}).
The analysis that follows~\cite{cg}, as well as that contained in the next section,
are essentially the same as that performed for the zeta-urn model~\cite{zeta1,zeta2}.

\subsection{Dynamics on the complete graph}

We wish to determine the temporal evolution of the occupation
probability $f_{k}(t)$.
Conservation of probability and of density yields 
\begin{eqnarray}
\sum_{k=0}^{\infty }f_{k}(t) &=&1, \label{sumrule1} \\
\sum_{k=1}^{\infty }k\,f_{k}(t) &=&\rho , \label{sumrule2}
\end{eqnarray}
where we have taken the thermodynamic limit $N\rightarrow \infty
,M\rightarrow \infty $, with fixed density $\rho =N/M$.
We consider a system with Poissonian initial distribution of occupation
probabilities,
\[
f_{k}(0)=\e^{-\rho }\frac{\rho ^{k}}{k!},
\]
i.e., such that initially particles are distributed at random amongst sites.

Since the equations (\ref{master}) are non linear they have no
explicit solution in closed form.
Yet one can extract
from them an analytical description of the dynamics of the system
at long times, both in the condensed phase, and at criticality.
The structure of the reasoning
borrows to former studies on urn models \cite{zeta1,zeta2}.
(For a review, see \cite{barcurn}.)

As we show below, there exists two different regimes in the evolution of the
system, both in the condensed phase or at criticality, which we study
successively.

\subsection*{(a) Nonequilibrium dynamics of condensation ($\rho >\rho _{c}$)}

Since $\bar{u}_{\eqq}=1$, we set, for large times,
\begin{equation}
\bar{u}_{t}\approx 1+A\,\,\varepsilon _{t},
\label{dbart}
\end{equation}
where the small time scale $\varepsilon _{t}$ is to be determined, and $A$
is an unknown amplitude.

\paragraph{Regime I: $k$ fixed, $t$ large.}

For $t$ large enough, sites empty ($u_k$) faster than they fill ($\bar{u}%
_{t}$). In this regime there is convergence to equilibrium, hence we set
\begin{equation}
f_{k}(t)\approx f_{k,\eqq}(1+v_{k}\,\varepsilon _{t}), \label{fkreg1}
\end{equation}
with $f_{k,\eqq}$ given by (\ref{fkc}), and where the $v_{k}$ are
unknown. This expression carried into (\ref{master}) yields the stationary
equation $\dot{f}_{k}=0$, because the derivative $\dot{f}_{k}$, proportional
to $\dot{\varepsilon}_{t}$, is negligible compared to the right-hand side. We
thus obtain an equation similar to the detailed balance condition:
\[
\frac{f_{k+1,\eqq}}{f_{k,\eqq}}\frac{1+v_{k+1}\,\varepsilon _{t}}{%
1+v_{k}\,\varepsilon _{t}}=\frac{1+A\,\varepsilon _{t}}{u_{k+1}}.
\]
Using (\ref{fkc}) and (\ref{pk}), we obtain, at leading order in $%
\,\varepsilon _{t}$, $v_{k+1}-v_{k}=A$, and finally
\begin{equation}
v_{k}=v_{0}+k\,A. \label{vk}
\end{equation}
At this stage, $v_{0}$ and the amplitude $A$ are still to be determined.

\paragraph{Regime II: $k$ and $t$ are simultaneously large.}

This is the scaling regime, with scaling variable $x=k\,\varepsilon _{t}$.
Following the treatment of \cite{zeta1,zeta2},
we look for a similarity solution of (\ref{master}) of the form
\begin{equation}
f_{k}(t)\approx (\rho -\rho _{c})\,\varepsilon _{t}^{2}\,g(x). \label{fkscalcond}
\end{equation}
We thus obtain for $g(x)$ the linear
differential equation
\[
g^{\prime \prime }(x)+\left( \frac{x}{2}-A+\frac{b}{x}\right) g^{\prime
}(x)+\left( 1-\frac{b}{x^{2}}\right) g(x)=0,
\]
with $\varepsilon_{t}\approx t^{-\frac{1}{2}}$.
This is precisely the differential equation found in \cite{zeta1,zeta2},
for the zeta-urn
model.
The amplitude $A$ can be determined by the fact that the equation has an
acceptable solution $g(x)$ vanishing as $x\rightarrow 0$ and $x\rightarrow
\infty $ \cite{zeta1}. 
The amplitude $A$ and the scaling function $g(x)$ are universal quantities, 
only depending on the
value of $b$. 
The sum rules 
(\ref{sumrule1}) and (\ref{sumrule2}) yield respectively 
\bea
\int_{0}^{\infty }\d x\,g(x)&=&\frac{v_{0}+A\rho_c}{\rho_c-\rho},\\
\int_{0}^{\infty }\d x\,x g(x)&=&1. 
\eea
The differential equation above has no closed form solution. However further
information on the form of the solution $g(x)$ can be found in \cite
{zeta1,zeta2}. 

An intuitive description of the dynamics of condensation
in the scaling regime is as follows.
The typical occupancy $k_{\rm cond}$ of the sites making
the condensate scales as $t^{\frac{1}{2}}$.
The total number of
particles in the condensate is equal to $M(\rho -\rho _{c})$,
the remaining $M\,\rho _{c}$ lying in the fluid.
Therefore the number of sites belonging
to the condensate scales as
$M (\rho -\rho _{c})t^{-\frac{1}{2}}$.

\subsection*{(b) Nonequilibrium critical dynamics ($\rho =\rho _{c}$)}

The analysis follows closely that done in \cite{zeta2}. 
We set
\[
\bar{u}_{t}\approx 1+A\,\varepsilon _{t} ,
\]
with $\varepsilon _{t}=t^{-\omega }$, where the exponent $\omega $ is to be
determined, and we consider the same two regimes as above.
In regime I, we still set (\ref{fkreg1}) for $f_{k}(t)$.
The reasoning
leading to the relationship $v_{k}=v_{0}+k\,A$ (see (\ref{vk}))
is still valid here.
In regime II, we look for a similarity solution to (\ref{master}) of the
form
\begin{equation}
f_{k}(t)\approx f_{k,\eqq}\,g_c(x)\qquad x=k\,t^{-\frac{1}{2}}.
\label{fkscal_crit}
\end{equation}
Indeed, for any large but finite time~$t$, the system looks critical, i.e.,
the occupation probabilities $f_k(t)$ have essentially converged toward
their equilibrium values~(\ref{fkc}), for $k\ll t^{1/2}$,
while for $k\gg t^{1/2}$ the system still looks disordered.
The $f_k(t)$ are expected to fall off very fast, which is confirmed
by the following analysis.

The sum rules (\ref{sumrule1}) and (\ref{sumrule2}) lead respectively to
the following equations, provided that $b>3$,
\begin{eqnarray}
v_{0}+A\rho _{c} &=&0, \label{v0_plus_aroc} \\
t^{-\omega }\left( v_{0}\rho _{c}+A\mu _{c}\right)
&=&t^{-(b-2)/2}(b-1)\Gamma (b)\int_{0}^{\infty }\d u\,u^{1-b}(1-g_c(u)),
\label{v0roc}
\end{eqnarray}
where $\mu _{c}=\sum
k^{2}\,f_{k,\eqq}$ is given in eq.~(\ref{second}).
Equation (\ref{v0roc}) fixes the value of $%
\omega $:
\begin{equation}
\omega =(b-2)/2. \label{omega}
\end{equation}
The differential equation obeyed by $g_c(x)$ is obtained by carrying (\ref
{fkscal_crit}) into (\ref{master}).
It reads
\[
g_c^{\prime \prime }(x)+\left( \frac{x}{2}-\frac{b}{x}\right) g_c^{\prime
}(x)=0,
\]
the solution of which is, with $g_c(0)=1$,
\begin{equation}
g_c(x)=\frac{2^{-b}}{\Gamma (\frac{b+1}{2})}\int_{x}^{\infty
}\d y\,y^{b}\e^{-y^{2}/4}.
\label{gu_anal}
\end{equation}
The fall-off of $g_c(x)$
for $x\gg 1$ is very fast: $g_c(x)\sim \exp (-x^{2}/4)$,
hence $f_{k}(t)\sim \exp (-k^{2}/4t)$.
We finally obtain
\[
A=\frac{(b-1)\Gamma (b)}{\mu _{c}-\rho _{c}^{2}}\int_{0}^{\infty
}\d u\,u^{1-b}(1-g_c(u))
=\frac{(b-2)(b-3)}{b-1}\Gamma \left( \frac{b}{2}\right)
.
\]

Let us mention that for any hopping rate of the form $u_k\approx 1+b/k$, the
scaling functions, $g(x)$ in the condensed phase (more
precisely: $g(x)/(\rho -\rho _{c})$),
and $g_c(x)$ at criticality, are universal.
In both cases the scaling variable is $x=k\,t^{-1/2}$.
The critical density $\rho _{c}$,
and, as a consequence, any quantity depending on $\rho_c$, such as
the amplitude $v_{0}$,
are non universal,
with values depending on the precise definition of
$u_k$.
As noted above, the amplitude $A$ is a universal quantity in the
condensed phase.

\subsection{Late stages of the dynamics and the case of one dimension}
\label{late}


As mentioned above, in the first stage of the dynamics, 
in the MF geometry, 
the number of most populated sites
decays as $M/t^{1/2}$.
Hence,
after a time of order $M^2$, the system contains a finite number
of highly populated sites, i.e., condensate precursors.

The late stage of the non-stationary dynamics,
where all but one of the precursors die out,
is thus expected to also last a length of time of the order of the diffusive timescale $M^2$.
This is substantiated by numerical simulations in~\cite{gross}.
Another argument is presented in section~\ref{last}.
The whole non-stationary process of the formation of the condensate
is therefore characterised by a single timescale 
\be
\tau_\noneq\sim M^2.
\ee

The same results hold for the 1DAS case.
The analysis relies upon numerical work or heuristic and scaling arguments~\cite{cg,gross}.

A similar scenario holds in the 1DS geometry,
the only difference being that $\tau_\noneq$ now scales as $M^3$.
The shift of the dynamical exponent by one unit in the 1DS geometry
has a common origin~\cite{cg,gross}: 
it stems from the Gambler's ruin problem~\cite{ruin}.
An analogous phenomenon is encountered for example in the coarsening law
for the domain growth, and
in the motion of a tagged particle, in 1D Kawasaki dynamics~\cite{kawa}.

We refer to the original references for further results
(scaling functions, critical case, etc.).

\section{Zero-range processes: nonequilibrium dynamics (II)}
\label{dyn2}

So far we considered the dynamics of one-time quantities, related to
the random variable $N_1(t)$.
We now explore another facet of the nonequilibrium dynamics of 
the ZRP with hopping rate $u_k=1+b/k$, namely the two-time
nonstationary aspects of its dynamics.
This essentially means that any function of the two times
depends on both times, instead of depending on their difference,
which would be the case at stationarity.
The situation here is analogous to that encountered when
a ferromagnetic spin system
is quenched from a high temperature, corresponding 
to an initial disordered configuration, to a lower temperature, $T\le T_c$
\cite{bray,barcferro}.

We consider the same ZRP as in the previous section, on the complete graph,
in the thermodynamic limit.
The system relaxes from a nonequilibrium initial
condition  towards equilibrium.
In order to characterize the fluctuations of 
the local density of particles,
$N_1(t)$, around its mean
$\mean{N_1(t)}=\rho$, we study
its associated two-time correlation and response functions,
and fluctuation-dissipation ratio.

\subsection{General framework}

The  connected two-time correlation function
of the density between time $s$ (waiting time) and time $t$ 
(observation time), with $s\le t$, is defined as
\be
C(t,s)=\mean{N_1(s)N_1(t)}-\rho^2.
\ee
It can be rewritten as
\be
C(t,s)=\sum_{k\ge1}k\,\g_k(t,s)-\rho^2,
\ee
where the function $\g_k(t,s)$ is defined by
\be
\g_k(t,s)=\sum_{j\ge1}j\,f_j(s)\,\P\{N_1(t)=k\mid N_1(s)=j\}
\ee
with the initial value at $t=s$
\be
\g_k(s,s)=k\,f_k(s).
\ee
Its temporal evolution for $t\ge s$ is given
by the master equation~(\ref{master}):
\beqa
\frad{\dpar\g_k(t,s)}{\dpar t}
&=&\mu_{k+1}\,\g_{k+1}+\lam_{k-1}\,\g_{k-1}
-\big(\mu_k+\lam_k\big)\g_k
\qquad (k\ge1),\nonumber
\\
\frad{\dpar\g_0(t,s)}{\dpar t}
&=&\mu_1\,\g_1-\lam_0\,\g_0.
\label{mdg}
\eeqa
The rates $\lam_k$ and $\mu_k$ are defined in~(\ref{rates}).
The rate $\lam_k$ only depends on the $f_k(t)$,
hence~(\ref{mdg}) are linear equations for the~$\g_k(t,s)$.

The local response function measures the influence on the mean density
on site number~1 of a perturbation in the canonically conjugate variable,
i.e., the local chemical potential acting on the same site.
Suppose that site number~1 is subjected to a small time-dependent
chemical potential $\alpha_1(t)$,
so that the total reduced energy of the system (see section~\ref{stat1}) is now
\be
\beta E\big(\ens\big)=\sum_{i=1}^M \beta E(N_i)+\alpha_1(t)N_1.
\ee
The mean density on site number~1 reads
\be
\mean{N_1(t)}=\rho+\int_0^t \d s\, R(t,s)\,\alpha_1(s)+\cdots,
\ee
where only the term linear in $\alpha(s)$ is written explicitly.
The kernel of the linear response is the two-time response function
\be
R(t,s)=\frac{\delta\mean{N_1(t)}}{\delta\alpha_1(s)}.
\ee
The temporal evolution of this function is given by a master equation
similar to~(\ref{master})~\cite{zeta2}.


The zero-range processes that we consider here
have a fast con\-ver\-gen\-ce towards equilibrium,
with a finite relaxation time $\tau_\rel$
in their fluid phase,
as is the case for a generic statistical-mechanical model
in its high-temperature disordered phase.
If the earlier time exceeds the relaxation time
($s\gg\tau_\rel$), the system is at equil\-ibri\-um.
One-time quantities take their equilibrium values.
Two-time quan\-ti\-ties, such as the correlation and response functions,
are invariant under time translations: 
\beq
C(t,s)=C_\eqq(\tau),\quad R(t,s)=R_\eqq(\tau),
\label{tti}
\eeq
where $\tau=t-s\ge0$.
They are related by the \fd theorem
\beq
R_\eqq(\tau)=-\frac{\d C_\eqq(\tau)}{\d\tau}.
\label{fdt}
\eeq

In the condensed phase and at criticality the relaxation time $\tau_\rel$
becomes infinite.
If the waiting time $s$ and the observation time $t$
are much smaller than $\tau_\rel$,
both time-translation invariance~(\ref{tti})
and the \fd theorem~(\ref{fdt}) are violated.
It is convenient~\cite{cuku}
to characterize departure from equilibrium by the \fd ratio
\beq
X(t,s)=\frac{R(t,s)}{\frad{\dpar C(t,s)}{\dpar s}}.
\label{xdef}
\eeq
In general, this dimensionless quantity depends on both times $s$ and $t$
and on the observable under consideration.
It may also exhibit a non-trivial scaling behavior in the two-time plane.
In all known cases it is observed that
\be
0\le X(t,s)\le1.
\ee

\subsection{Application: ZRP with condensation ($u_k=1+b/k$)}


\subsubsection*{Nonequilibrium critical dynamics $(\rho=\rho_c)$.}

Let us first note that the variance of the population of site number~1
converges to its equilibrium value $C_\eqq=\mu_c-\rho_c^2$ as a power law:
\beq
C(t,t)=\mean{N_1(t)^2}-\rho_c^2\approx
C_\eqq-\frac{2^{3-b}\,t^{-(b-3)/2}}
{(b-3)\,\Gamma\left((b+1)/2\right)}.
\label{cvar}
\eeq
The derivation of the behaviour of the two-time density correlation and response functions
is the same as in~\cite{zeta2}.
In the nonequilibrium scaling regime ($s,t\gg1$),
one finds
\beq
\matrix{
C(t,s)\approx s^{-(b-3)/2}\,\Phi(x),\hfill\cr
{\hskip -3.05truemm}
\frad{\dpar C(t,s)}{\dpar s}\approx s^{-(b-1)/2}\,\Phi_1(x),\hfill\cr
R(t,s)\approx s^{-(b-1)/2}\,\Phi_2(x),\hfill
}
\label{csca}
\eeq
where 
\be
x=t/s\ge1.
\ee
As a consequence, 
in the scaling regime,
the \fd ratio $X(t,s)$ only depends on $x$:
\be
X(t,s)\approx\X(x)=\frac{\Phi_2(x)}{\Phi_1(x)}.
\ee
The dimensionless scaling function $\X(x)$ is universal,
and it admits a non-trivial limit value in the regime where the two
time variables $s$ and $t$ are well separated in the scaling regime~\cite{glcrit}:
\be
X_\infty=\lim_{s\to\infty}\lim_{t\to\infty}X(t,s)=\X(\infty).
\ee

Explicit expressions for the above scaling functions can be derived,
using a spectral decomposition in Laguerre polynomials~\cite{zeta2}.
The limit \fd ratio thus obtained
\be
X_\infty=\frac{b+1}{b+2}\quad(b>3),
\ee
lies in an unusually high range ($4/5<X_\infty<1$) for a critical system.
Indeed, statistical-mechanical models such as ferromagnets
are observed to have $0<X_\infty\le1/2$ at their critical point.
The upper bound $X_\infty=1/2$, corresponding to the mean-field
situation~\cite{glcrit},
is also observed in a range of simpler models~\cite{cuku,glglau}.

The above results illustrate general predictions
on nonequilibrium critical dy\-na\-mics~\cite{janssen,glglau,glcrit,barcferro}.
The exponent of the waiting time~$s$ in the first line of~(\ref{csca})
already appears in~(\ref{cvar}).
It is related to the anomalous dimension of the observable under consideration,
and would read $(d-2+\eta)/z_c$ for a $d$-dimensional ferromagnet,
where $\eta$ is the equilibrium correlation exponent
and $z_c$ the dynamical critical exponent.
The scaling functions $\Phi(x)$, $\Phi_{1,2}(x)$
are universal up to an overall multiplicative constant,
and they obey a common power-law fall-off in $x^{-b/2}$.
The latter exponent is not related to exponents pertaining to
usual equilibrium critical dynamics.
It reads $-\lambda_c/z_c=\Theta_c-d/z_c$ for a ferromagnet,
where $\lambda_c$ is the critical autocorrelation exponent~\cite{huse}
and $\Theta_c$ is the critical initial-slip exponent~\cite{janssen}.

\subsubsection*{Nonequilibrium dynamics of condensation  $(\rho>\rho_c)$.}

\begin{figure}[htb]
\begin{center}
\includegraphics[angle=90,width=.8\linewidth]{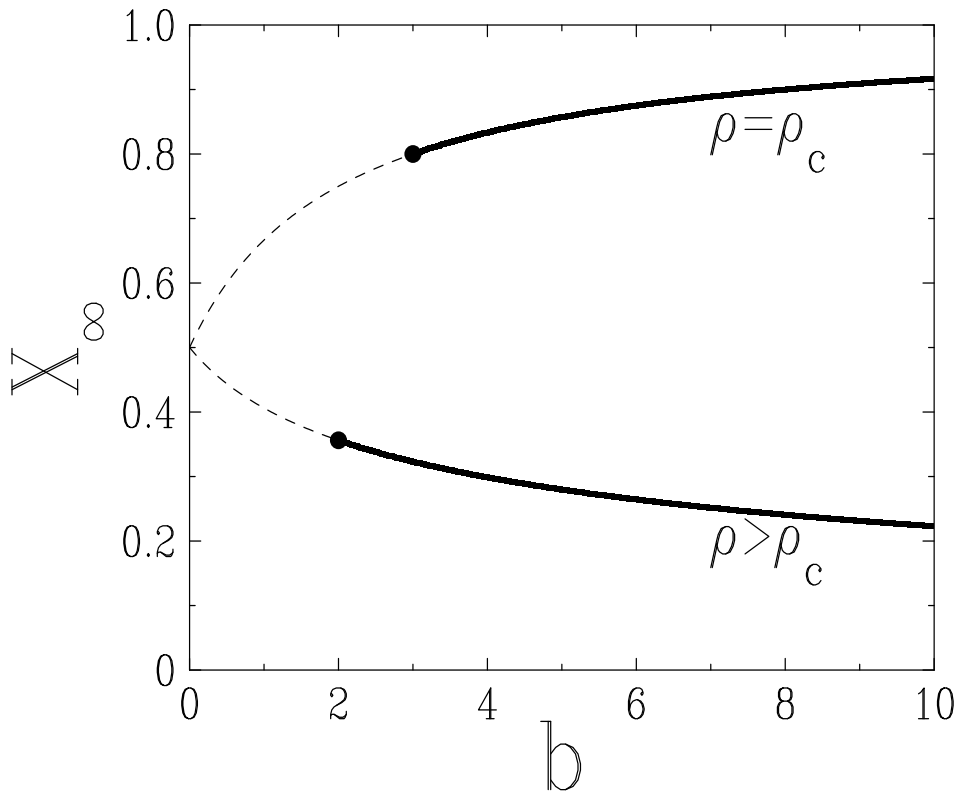}
\caption{\small
Plot of the limit \fd ratio $X_\infty$ against  $b$.
Upper curve: critical point ($b>3$, $\rho=\rho_c$).
Lower curve: condensed phase ($b>2$, $\rho>\rho_c$).
Thin dashed lines: continuation of the results to lower values of $b$.}
\label{fx}
\end{center}
\end{figure}

In the scaling regime, 
two-time quantities
are found to scale as~\cite{zeta2} 
\beq
\matrix{
C(t,s)\approx(\rho-\rho_c)s^{1/2}\,\Phi(x),\qquad(x=t/s),\hfill\cr
{\hskip -3.truemm}
\frad{\dpar C(t,s)}{\dpar s}\approx(\rho-\rho_c)s^{-1/2}\,\Phi_1(x),\hfill\cr
R(t,s)\approx(\rho-\rho_c)s^{-1/2}\,\Phi_2(x),\hfill\cr
{\hskip -.5truemm}X(t,s)\approx\X(x)=\frad{\Phi_2(x)}{\Phi_1(x)}.\hfill
}
\label{cosca}
\eeq
The scaling functions $\Phi(x)$, $\Phi_{1,2}(x)$
have finite values, both at coinciding times $(x=1)$ and in the limit
of large time separations $(x=\infty)$.
The limit \fd ratio $X_\infty=\X(\infty)$
depends continuously on $b$ throughout the condensed phase
($b>2$), and vanishes only as
\be
X_\infty=b^{-1/2}-\frac{b^{-3/4}}{4}+\cdots
\ee
for $b$ large, which corresponds formally to low temperature, 
while coarsening systems are known~\cite{xzero}
to have identically $X_\infty=0$ throughout their low-temperature phase.
In figure~\ref{fx} a summary of the values of $X_\infty$ is presented.

This dynamics is different from the usual phase-ordering dynamics~\cite{bray}.
Indeed, when a ferromagnet
is quenched below its critical temperature,
domain growth and phase separation
take place in a statistically homogeneous way, at least for an infinite system.
In the present situation,
condensation takes place in a very inhomogeneous fashion,
since fewer and fewer sites are involved in the process.

\subsection{One dimension}

For both the symmetric and asymmetric cases the response can
be defined in the same fashion as above. There is no analytical
tools at our disposal to compute these functions, even in the scaling regime.
However, for the symmetric case, the fluctuation-dissipation still holds
at equilibrium, while it should be violated in the stationary state of
the asymmetric case.

\section{Stationary dynamics of the condensate}
\label{dyn3}

\subsection{The question posed}

\begin{figure}[htb]
\begin{center}
\includegraphics[angle=90,width=.6\linewidth]{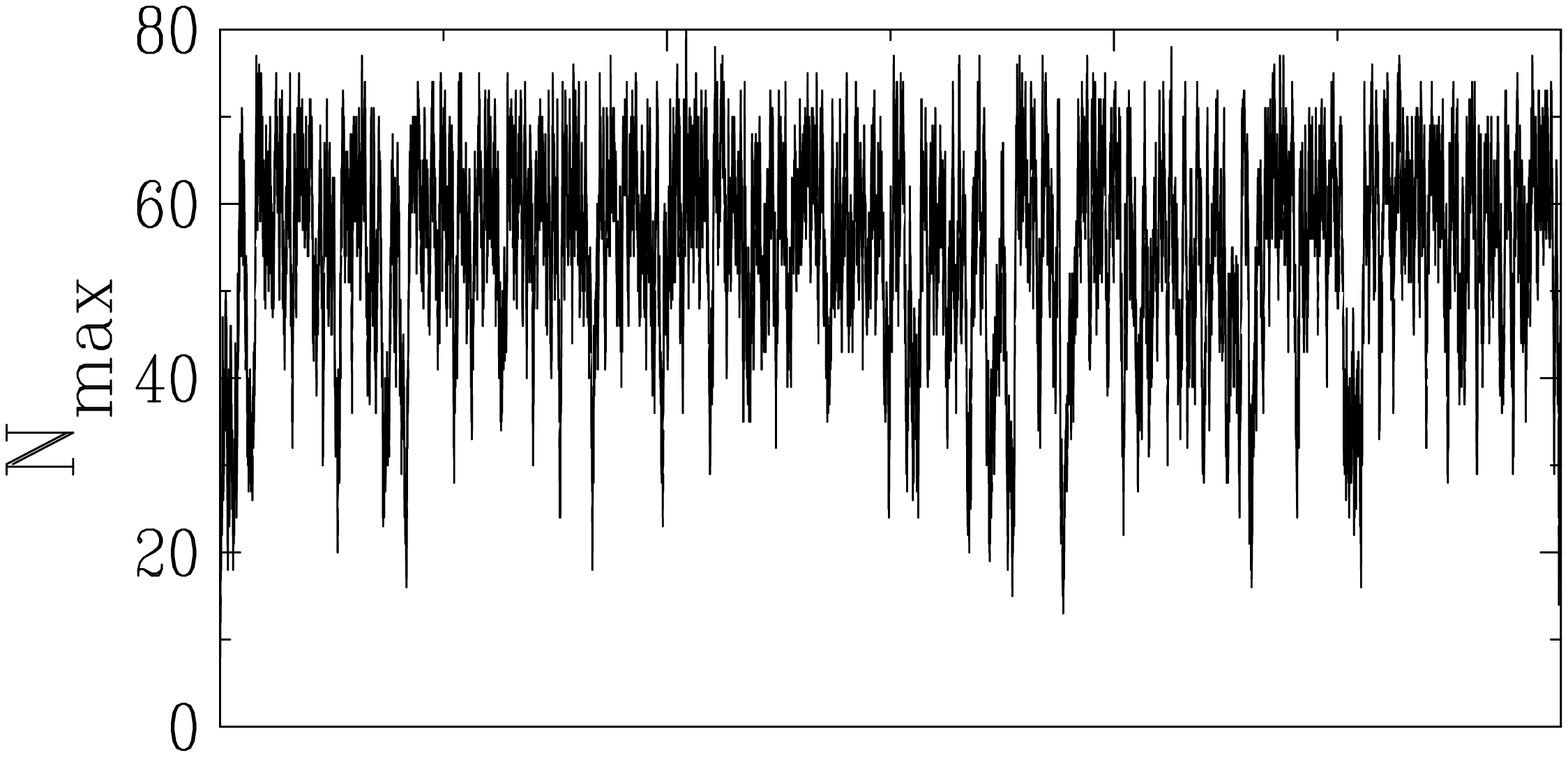}
\vskip -9truemm
\includegraphics[angle=90,width=.6\linewidth]{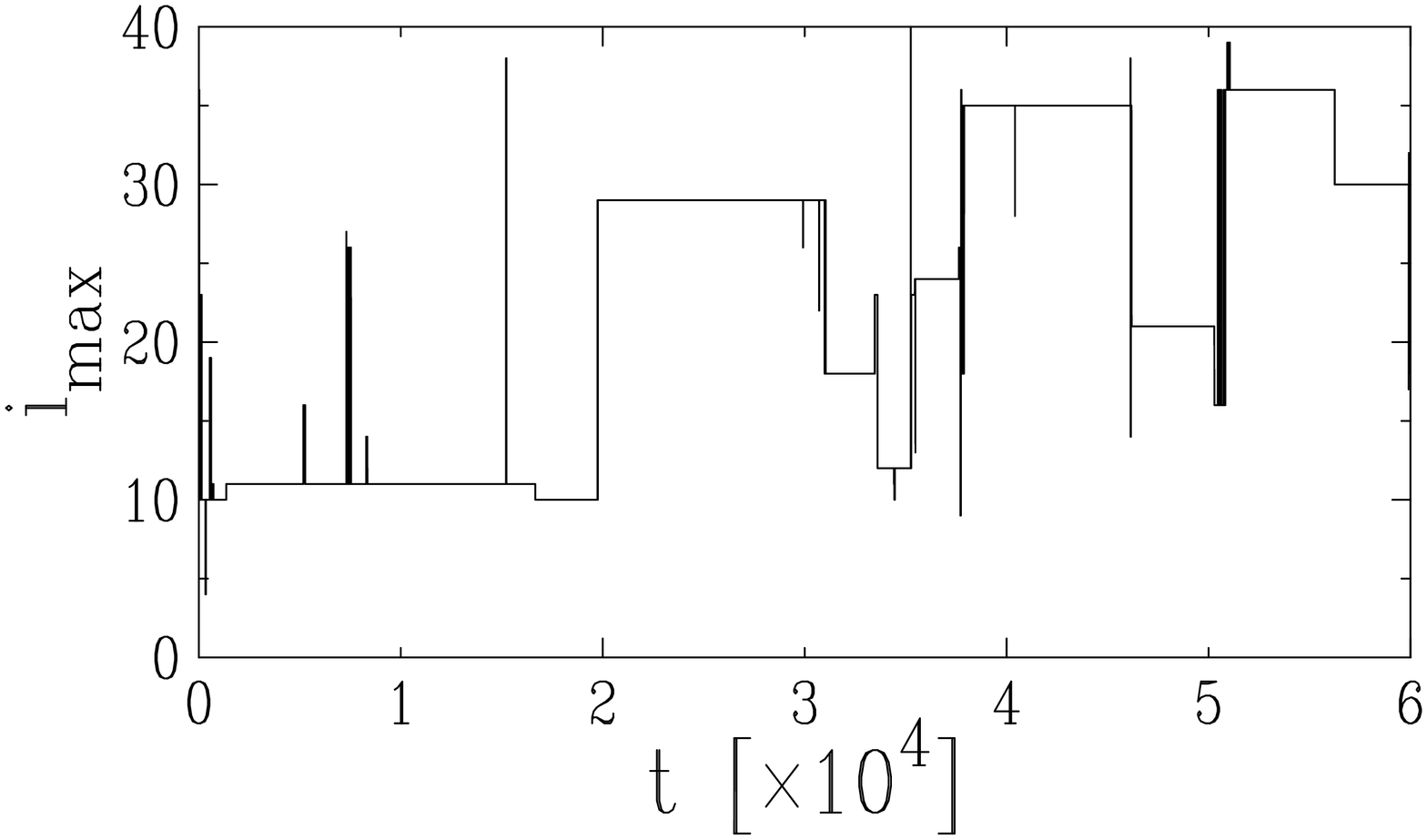}
\caption{\small
Dynamics of the condensate (1DS geometry with $b=4$, $M=40$, $N=80$).
Upper panel: instantaneous number of particles
$N_\max(t)$ on the most populated site.
Lower panel: location $i_\max(t)$ of that site.}
\label{f1}
\end{center}
\end{figure}

Consider a ferromagnetic system,
an Ising spin system for instance. 
At equilibrium in the low temperature
phase, the spin symmetry is spontaneously broken.
There are two possible equilibrium states, one with positive magnetisation,
the other one with negative magnetisation.
However, if one observes a large but finite system, then
as time passes, 
the magnetisation keeps changing sign,
the system flipping between the two possible equilibrium states.
Ergodicity is restored for a finite system.
The typical time between two flips
is exponential in $L^{d-1}$, where $L$ is the linear size of the system,
and $d$ the dimension of space.

A similar situation occurs for in the condensed phase of a ZRP.
Here the spontaneously broken symmetry 
is translational invariance. 
For a large but finite system in the stationary state, as time passes,
the condensate keeps moving across the system.
It spends long lengths of time on a given site,
before suddenly disappearing and reappearing on another site.
The typical value of these lengths of time defines the characteristic time
$\tau$ of the dynamics of the condensate.
The aim of this section is to analyse the nature of this motion and
in particular to characterise
how $\tau$ scales with the system size $M$.

\subsection{Numerical observations}

An intuitive understanding of the phenomenon is easily
gained by performing Monte-Carlo simulations.
These simulations, done
in the three geometries: mean-field (MF),
one-dimensional asymmetric (1DAS) ($p=1$),
and one-dimensional symmetric (1DS) ($p=q=1/2$), lead to a common picture.

The condensate is immobile for rather long lapses of time;
it then performs sudden random non-local jumps all over the system,
at Poissonian times whose characteristic scale
grows rapidly with the system size $M$.
Figure~\ref{f1} illustrates this process for the 1DS case,
for a system of size $M=40$, with $N=80$ particles, i.e., $\rho=2$,
and $b=4$, hence $\rho_c=1/2$.
The upper panel shows the track of the instantaneous number of particles
$N_\max(t)$ on the most populated site.
The signal for $N_\max(t)$ fluctuates around 
$\Delta\equiv M(\rho-\rho_c)=60$,
the mean size of the condensate.
The lower panel shows the label $i_\max(t)$ of that site,
i.e., the location of the condensate.
The non-local character of the motion of the condensate is clearly visible,
whereas the longest lapses of time where the condensate stays still
give a heuristic measure of the characteristic time $\tau$.

We show in what follows that $\tau\sim M^b$ for the fully connected geometry 
and the directed case, while
$\tau\sim M^{b+1}$ for the symmetric case.
Moving the condensate is therefore slower than forming the condensate 
($\tau_\noneq\sim M^2$, $\tau_\noneq\sim M^3$ respectively, see section~\ref{late}).

\subsection{Theoretical analysis}

All the idea relies on a problem of barrier crossing.
Defining the potential as $V_k=-\ln f_{k,\eq}$,
then a dip in the probability $f_{k,\eq}$ corresponds to
a barrier in the potential.
The flipping time $\tau$ is the time to cross the barrier, or the first-passage
time from right to left.
Let us explain these ideas in more detail.

\begin{figure}[htb]
\begin{center}
\includegraphics[angle=90,width=.6\linewidth]{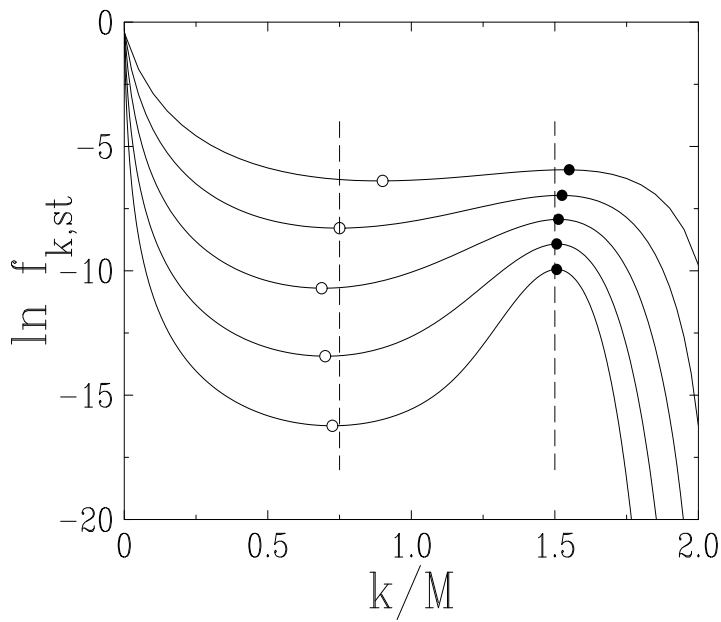}
\caption{\small
Logarithmic plot of the occupation probability $f_{k,\eq}$
in the condensed phase ($b=4$, $\rho=4\rho_c=2$), against the ratio $k/M$.
Top to bottom: $M=20$, 40, 80, 160, and 320.
Full lines: $f_{k,\eq}$ obtained by~(\ref{zrecrel}) and~(\ref{fkeq}).
Full (empty) symbols: maxima (minima) of occupation probability.
Dashed vertical lines:
asymptotic locations of the minima: $k/M=\D/(2M)=(\rho-\rho_c)/2=3/4$,
and of the maxima: $k/M=\D/M=\rho-\rho_c=3/2$.}
\label{ffk}
\end{center}
\end{figure}

We first analyse the behaviour of the occupation
probability $f_{k,\eq}$ in the condensed phase.
Figure~\ref{ffk} shows a logarithmic plot of 
$f_{k,\eq}$,
computed using equations~(\ref{zrecrel}) and~(\ref{fkeq}),
against the ratio $k/M$, for $b=4$, $\rho=2$, and several values of $M$.
This plot exhibits the following features.
\begin{itemize}
\item
For $k\ll M$, the distribution~$f_{k,\eq}$ is approximately given
by the power law~(\ref{fkc}) of an infinite critical system.
\item
The contribution of the condensate slowly builds up as a probability hump
around $k=\D=M(\rho-\rho_c)$, the mean number of excess particles.
\item
One observes a broad and shallow probability ``dip'' in the region located
between the critical background and the condensate hump, i.e.,
in the region $k\gg1$ and $\D-k\gg1$.
\end{itemize}
The region of the dip is dominated by
configurations
where the excess particles are shared by {\it two} sites.
Indeed, one has (see~\cite{gl2005} for a proof)
\beq
f_{k,\eq}\approx(b-1)\Gamma(b)\frac{\D^b}{k^b(\D-k)^b}
\qquad ( k\gg1, \D-k\gg1).
\label{fkdip}
\eeq
The observed locations of the maxima ($k\approx\Delta$) and minima
($k\approx\Delta/2$) of the occupation probability
corroborate this picture, as explained in the caption of figure~\ref{ffk}.

These observations lead to the following crude estimate
for the characteristic time:
\beq
\tau\sim\frac{1}{f_\min},
\label{arrmin}
\eeq
since the minimum $f_\min$ of  $f_{k,\eq}$
corresponds to a barrier to cross, in the spirit of the Arrhenius law.
The limiting scale of time is that required for the passage of this potential barrier.
Eq.~(\ref{fkdip})
implies that $f_\min$ is reached near the middle of the dip region
$(k\approx N/2)$, and therefore~(\ref{arrmin}) yields
\beq
\tau\sim\D^b.
\label{arrdip}
\eeq

We now present a more precise treatment.
Assume that the condensate is on site number 1
at the initial observation time ($t=0$).
The number $N_1(t)$ of particles on that site is initially
very large, $N_1(0)\approx\D$, and therefore evolves slowly,
until the condensate dissolves into the critical background.
Thus 
\begin{itemize}
\item
We single out $N_1(t)$ as the collective co-ordinate of the system,
that is the appropriate slow variable describing the dynamics of the condensate.
\item
We model the dynamics of $N_1(t)$ by~(\ref{2sites}),
i.e., by a biased diffusive motion on the interval $k=0,\dots,N$.
The left hopping rate is taken equal to the microscopic rate: $\mu_k=u_k$.
The right hopping rate $\la_k$ is chosen such that, in the stationary state,
the probability $f_{k,\eq}$ of the effective model coincide with the
occupation probability~(\ref{fkeq}) of the original ZRP.
The detailed balance condition~(\ref{db2}) yields
\be
\la_k=\frac{\mu_{k+1}f_{k+1,\eq}}{f_{k,\eq}}
=\frac{Z_{M-1,N-k-1}}{Z_{M-1,N-k}}
=\bar u_\eq(M-1,N-k),
\ee
where the right side of the equation is the average rate coming
from $M-1$ sites, containing $N-k$ particles (see (\ref{ubarstat})).
The rate $\la_k$ is thus a function of $k$, $M$, and $N$.
For $M=2$, this formula gives $\la_k=u_{N-k}$, as expected.
In the fluid phase, in the thermodynamic limit,
the rates~$\la_k$ converge to $z_0$, defined in~(\ref{col}).
Finally, for the condensed phase, in the dip region, we obtain
\be
\la_k\approx 1+\frac{b}{\D-k}\equiv u_{\D-k}.
\ee
\end{itemize}

This effective description reduces the full model to a
Markovian model for one degree of freedom in an {\em asymmetric} potential.
The two valleys of the potential are separated by a high (power-law) barrier.
The left potential valley, corresponding to the critical background,
has a weight $P_\L\approx1$, whereas the right potential valley,
corresponding to the hump of the condensate, has a weight $P_\R\approx1/M\ll1$
(see section~\ref{canon}).

In this framework the stationary dynamics
of the condensate is characterised by a single diverging timescale.
We choose to define this timescale, denoted by~$\tau_\ma$,
to be the crossing time~$T_\L$ from the right valley to the left one
in the effective Markovian problem.
The characteristic time is thus expressed by
\beq
\tau_\ma\equiv T_\L=\sum_{\ell=1}^N\frac{1}{\mu_\ell f_{\ell,\eq}}
\sum_{m=\ell}^Nf_{m,\eq},
\label{tauma}
\eeq
in terms of known quantities, the rates $\mu_k$
and the stationary probabilities $f_{k,\eq}$.
Its asymptotic growth is easily determined by noting that~(\ref{tauma})
is dominated by the behaviour of the probability $f_{k,\eq}$ in the region of
the dip.
Hence, inserting the expression~(\ref{fkdip}) into~(\ref{tauma}),
and evaluating the sum as an integral, we obtain
\beq
\tau_\ma\approx\frac{b\Gamma(b+1)}{(b-1)\Gamma(2b+2)}\,\frac{\D^{b+1}}{M}
=\frac{b\Gamma(b+1)}{(b-1)\Gamma(2b+2)}\,(\rho-\rho_c)^{b+1}M^b.
\label{zrpt1}
\eeq

In order to compare the above theoretical predictions to the measured 
flipping time $\tau$, we compute
the two-time stationary correlation function 
\be
C(t,0)=\mean{N_1(t)N_1(0)}-\rho^2.
\ee
This quantity decays exponentially with a relaxation constant which
gives a natural measure of $\tau$.
It is found that $\tau\sim\tau_\ma\sim M^b$ in the MF and 1DAS geometries,
and that $\tau\sim M\tau_\ma\sim M^{b+1}$ in the 1DS geometry.
For the latter case the occurence of one supplementary power 
in the system size has the same
origin as for nonequilibrium dynamics.

\subsection{Last remarks}
\label{last}

Table~\ref{one} summarises the values of the dynamical exponents $z$ and $Z$,
such that $\tau_\noneq\sim M^z$ and $\tau\sim M^Z$,
where $\tau$
is the characteristic timescale for the stationary motion of the condensate.
\begin{table}[ht]
\begin{center}
\begin{tabular}{|c|c|c|}
\hline
Geometry&$z$&$Z$\\
\hline
MF, 1DAS&2&$b$\\
\hline
1DS&3&$b+1$\\
\hline
\end{tabular}
\caption{\small Non-stationary and stationary dynamical exponents
of the ZRP with static exponent $b>2$.}
\label{one}
\end{center}
\end{table}
As recalled above, the non-stationary dynamical exponents
are insensitive to the exponent $b$, and more generally to the statics,
provided the system is in its condensed phase.
This feature is easily understood in the context of the
Markovian Ansatz proposed in the present work.
Indeed the last stage of the formation of the condensate,
i.e., the disap\-pearance of the smaller of the last two precursors,
implies no barrier crossing.
In terms of the occupation of the condensate,
it  corresponds to the transition from $N_1$ to $\D$,
where the initial occupation $N_1$ of the larger precursor
was already larger than $\D/2$,
corresponding to the top of the potential barrier.
This explains why $\tau_\noneq$ is given by the diffusive timescale,
both in the framework of the Markovian Ansatz
and in the MF and 1DAS geometries.

\section{Further references}
Complementary aspects to the present notes
can be found in~\cite{evans2,landim,jona,rakos}.

\ack
It is a pleasure to thank J-M Drouffe and J-M Luck, with whom
I have been collaborating through the years on the subject of these notes, as well as
R Blythe, M Evans, S Grosskinsky, T Hanney,
E Levine, S Majumdar, D Mukamel, G Sch\"utz, and H Spohn
for interesting discussions.

\appendix
\setcounter{equation}{0}
\def\theequation{A.\arabic{equation}}
\section*{Appendix: Proof of eqs.~(\ref{db}) and (\ref{constraint})} 

We recall the fundamental equation (\ref{fund})
\beqa\label{app:fund}
p(W_{k,l}-W_{k,0})+q(W_{l,k}-W_{l,0})=\nonumber\\
p\, \left(W_{k+1,l-1}\,\frac{p_{k+1} p_{l-1}}{p_k p_l}-W_{l,0}\right)
+q\,\left( W_{l+1,k-1}\,\frac{p_{k-1} p_{l+1}}{p_k p_l}-W_{k,0}\right).
\eeqa

We first prove (\ref{db})
for the symmetric case, $p=1/2$.
Setting $x_k=W_{k,l}\,p_kp_l$, where $k+l=n$,
(\ref{app:fund}) can be rewritten as
\be
x_k-x_{n-(k-1)}=x_{k+1}-x_{n-k}.
\ee
This expression is therefore a constant independent of $k$, which is
equal to zero, as can be seen by taking $k=n$. 
We thus obtain 
\be
x_{k+1}=x_{n-k}
\ee
which is 
the detailed balance condition
\beq\label{app:db}
p_k p_l\,W_{k,l}
=p_{k-1} p_{l+1}\,W_{l+1,k-1}.
\eeq

We now show that in the general case, $p\ne1/2$, eq.~(\ref{app:fund}) yields 
two constraints on the rate: eq.~(\ref{app:db}) to be interpreted as the pairwise balance condition, 
and eq. (\ref{constraint}) 
\beq\label{app:constraint}
W_{l,k}-W_{k,l}=W_{l,0}-W_{k,0}.
\eeq
Set $a_k=p_kp_l\,W_{k,l}$.
Eq.~(\ref{app:fund}) can be rewritten as
\beq\label{rec}
y_{k+1}-y_k=(p-q)(a_{n-k}-a_k)
\eeq
where
\beq\label{yk}
y_k=p\,x_k-q\,x_{n-(k-1)},\qquad y_0=0.
\eeq
If 
\beq\label{sym}
y_{k+1}=y_{n-k}
\eeq
then it follows immediately that
$x_{k+1}=x_{n-k}$,
which is the condition for pairwise balance seen above.
This relation itself plugged into (\ref{yk}), yields
\be
x_{k+1}-x_k=a_{n-k}-a_k
\ee
which is (\ref{app:constraint}).
In order to prove (\ref{sym}) we set
\be
A_k=a_1+\ldots+a_k
\ee
i.e. $a_k=A_k-A_{k-1}$.
We thus have
\be
y_k+(p-q)(A_{k-1}+A_{n-k})=y_{k+1}+(p-q)(A_{k}+A_{n-k-1})
\ee
which is equal to $(p-q)A_n$, hence
\be
y_k=(p-q)(A_n-A_{n-k}-A_k+a_k).
\ee
Therefore $y_k-(p-q)a_k$ is symmetric in the change $k\to n-k$, and finally
\be
y_k-(p-q)a_k=y_{n-k}-(p-q)a_{n-k}
\ee
which, using (\ref{rec}), yields (\ref{sym}).
%
%
\section*{References}

\end{document}